\documentclass[letterpaper, 10 pt, conference]{ieeeconf}  

\IEEEoverridecommandlockouts                              

\usepackage[T1]{fontenc}
\usepackage{cite}
\usepackage{amsmath,amssymb,amsfonts}
\usepackage{dsfont}

\usepackage{bm}
\usepackage{bbm}

\usepackage{algorithm}
\usepackage{algpseudocodex}
\renewcommand{\algorithmicrequire}{\textbf{Input:}}
\renewcommand{\algorithmicensure}{\textbf{Output:}}

\newcommand{\code}[1]{\texttt{#1}}
\algrenewcommand\algorithmicforall{\textbf{for each}}

\usepackage{graphicx}
\graphicspath{{./graphics/}}
\definecolor{plotly_blue}{RGB}{51,102,204}
\definecolor{plotly_green}{RGB}{0,128,0}
\usepackage{subcaption}
\usepackage{multirow}
\usepackage{siunitx}
\usepackage{verbatim}
\usepackage{hyperref}
\hypersetup{
    colorlinks=true,
    linkcolor=blue,
    filecolor=magenta,      
    urlcolor=blue,
}
\usepackage{todonotes}
\usepackage{definitions}
\usepackage[absolute,overlay]{textpos}

\title{\LARGE \bf
Cooperative Trajectory Planning in Uncertain Environments with Monte Carlo Tree Search and Risk Metrics}

\author{Philipp Stegmaier$^{1}$, Karl Kurzer$^{1}$ and J. Marius Z\"ollner$^{1}$
\thanks{$^{1}$Karlsruhe Institute of Technology, 76131 Karlsruhe, Germany\newline {\tt\small \{philipp.stegmaier, karl.kurzer, marius.zoellner\}@kit.edu}}%
}

\begin{document}
\begin{textblock*}{\textwidth}(19mm,10mm)
\footnotesize
\noindent \copyright 2022 IEEE.  Personal use of this material is permitted.  Permission from IEEE must be obtained for all other uses, in any current or future media, including reprinting/republishing this material for advertising or promotional purposes, creating new collective works, for resale or redistribution to servers or lists, or reuse of any copyrighted component of this work in other works.\\
\textit{2022 IEEE International Conference on Intelligent Transportation Systems (ITSC)}
\end{textblock*}

\maketitle
\thispagestyle{empty}
\pagestyle{empty}

\begin{abstract}
Automated vehicles require the ability to cooperate with humans for smooth integration into today's traffic. While the concept of cooperation is well known, developing a robust and efficient cooperative trajectory planning method is still a challenge. One aspect of this challenge is the uncertainty surrounding the state of the environment due to limited sensor accuracy. This uncertainty can be represented by a Partially Observable Markov Decision Process. Our work addresses this problem by extending an existing cooperative trajectory planning approach based on Monte Carlo Tree Search for continuous action spaces. It does so by explicitly modeling uncertainties in the form of a root belief state, from which start states for trees are sampled. After the trees have been constructed with Monte Carlo Tree Search, their results are aggregated into return distributions using kernel regression. We apply two risk metrics for the final selection, namely a Lower Confidence Bound and a Conditional Value at Risk. It can be demonstrated that the integration of risk metrics in the final selection policy consistently outperforms a baseline in uncertain environments, generating considerably safer trajectories.
\end{abstract}


\section{INTRODUCTION}
In the context of automated driving, uncertainties about the real state of the environment exist due to the limited accuracy of sensor systems. Further, the outcomes of actions are uncertain as they depend on the unknown intentions of other traffic participants as well as imprecise actuators and controllers. Hence, trajectory planning algorithms should incorporate these sources of uncertainty to reduce risks by yielding more robust trajectories.

While Partially Observable Markov Decision Processes (POMDPs) provide a sound mathematical framework for sequential decision making in uncertain environments, POMDP solutions are intractable for all but the smallest problems due to their inherent complexity \cite{Ross2008} (i.e., \textit{curse of dimensionality} and \textit{curse of history} \cite{Silver2010a}). Hence, approximation methods have been developed \cite{Klimenko2014} that are capable of determining local policies online \cite{Ross2008} if the state and action space are low dimensional. These methods are applied for interactive decision making in uncertain driving situations by a strong reduction of the state and action spaces \cite{Hubmann2017, Hubmann2018, Schoerner2019, Ulbrich2013}.

The cooperative trajectory planning algorithm this work is based on plans in a continuous state and action space for each agent \cite{Kurzer2018b}. Accounting for all possible states in the multi-agent setting results in a complete solution space. The arising complexity forbids the use of standard POMDP solution methods. Hence, we address the uncertainty by substituting the initial state $\state_0$ of a Markov Decision Process (MDP) with an initial belief $\belief_0$, modeling a distribution over all possible states according to the current observation \cite{Kaelbling1998}. All successor states are treated as fully observable in the corresponding planning cycle, and thus standard MDP logic can be applied.

Our main contribution is the extension of prior work \cite{Kurzer2018b} to be capable of handling uncertainties due to limited sensor accuracy. We combine determinization, Monte Carlo Tree Search (MCTS), and kernel regression to obtain return distributions. By applying risk metrics to the return distributions, robust actions are determined.

\begin{figure}
	\centering
	\def\svgwidth{0.65\columnwidth}
\begingroup%
  \makeatletter%
  \providecommand\color[2][]{%
    \errmessage{(Inkscape) Color is used for the text in Inkscape, but the package 'color.sty' is not loaded}%
    \renewcommand\color[2][]{}%
  }%
  \providecommand\transparent[1]{%
    \errmessage{(Inkscape) Transparency is used (non-zero) for the text in Inkscape, but the package 'transparent.sty' is not loaded}%
    \renewcommand\transparent[1]{}%
  }%
  \providecommand\rotatebox[2]{#2}%
  \newcommand*\fsize{\dimexpr\f@size pt\relax}%
  \newcommand*\lineheight[1]{\fontsize{\fsize}{#1\fsize}\selectfont}%
  \ifx\svgwidth\undefined%
    \setlength{\unitlength}{1388.97637795bp}%
    \ifx\svgscale\undefined%
      \relax%
    \else%
      \setlength{\unitlength}{\unitlength * \real{\svgscale}}%
    \fi%
  \else%
    \setlength{\unitlength}{\svgwidth}%
  \fi%
  \global\let\svgwidth\undefined%
  \global\let\svgscale\undefined%
  \makeatother%
  \begin{picture}(1,0.83920998)%
    \lineheight{1}%
    \setlength\tabcolsep{0pt}%
    \put(0,0){\includegraphics[width=\unitlength,page=1]{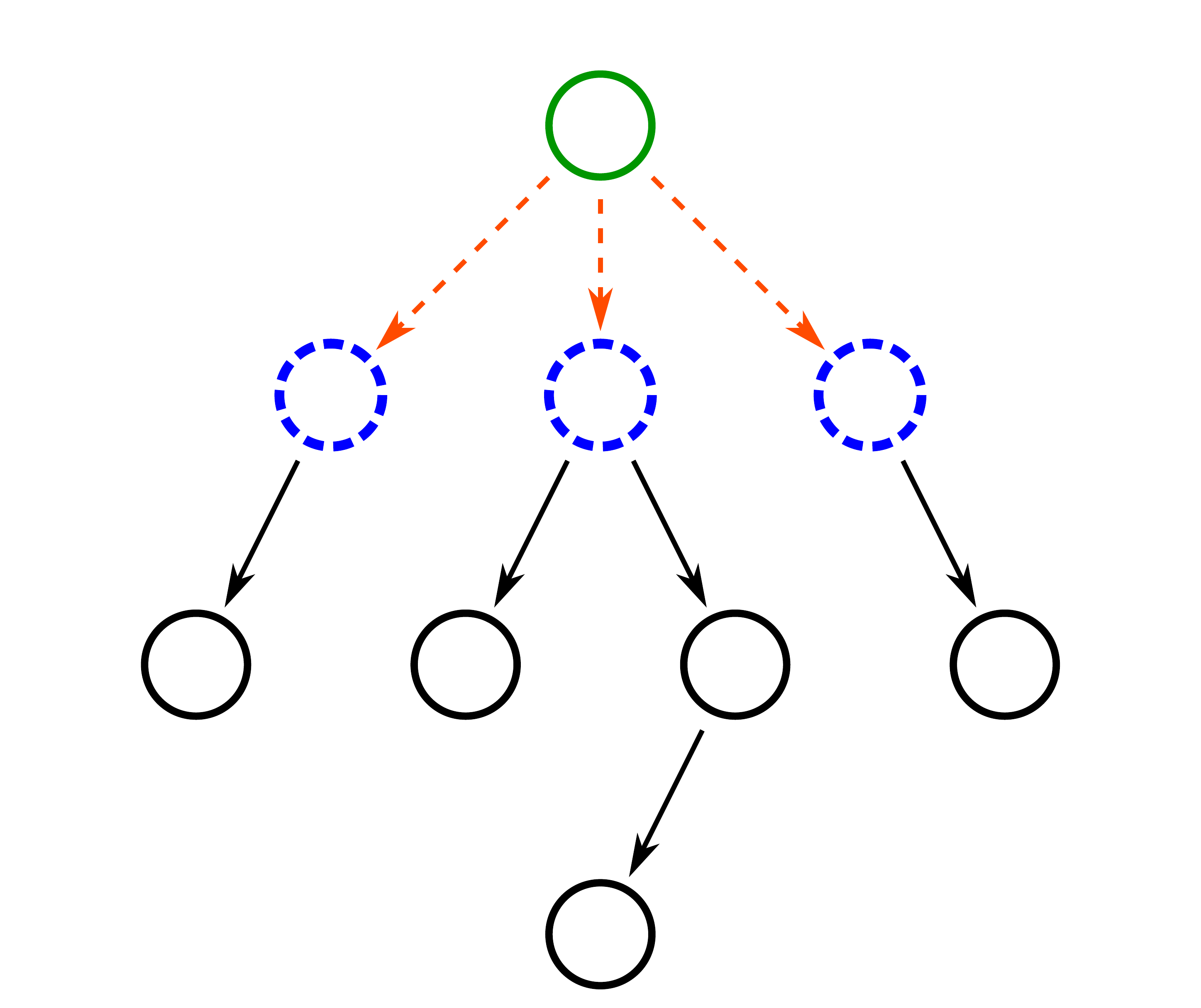}}%
    \put(0.55904494,0.7261513){\makebox(0,0)[lt]{\lineheight{1.25}\smash{\begin{tabular}[t]{l}$\belief_0$\end{tabular}}}}%
    \put(0.65445684,0.64207343){\makebox(0,0)[lt]{\lineheight{1.25}\smash{\begin{tabular}[t]{l}$\startState \sim \belief_0$\end{tabular}}}}%
    \put(0.78353475,0.50166151){\makebox(0,0)[lt]{\lineheight{1.25}\smash{\begin{tabular}[t]{l}$\startState \in \startStates$\end{tabular}}}}%
    \put(0.81076344,0.41758365){\makebox(0,0)[lt]{\lineheight{1.25}\smash{\begin{tabular}[t]{l}$\actionVec \in \actionSpace$\end{tabular}}}}%
    \put(0.89577968,0.27717173){\makebox(0,0)[lt]{\lineheight{1.25}\smash{\begin{tabular}[t]{l}$\state \in \states$\end{tabular}}}}%
  \end{picture}%
\endgroup%

	\caption{Using the distribution of the belief state $\belief_0$, possible start states are sampled. For each start state $\startState$, an MCTS is run to determine the action values $\actionValueRVar(\startState, \cdot)$, yielding a return distribution over start states $\startStates$.}
	\label{fig:overview}
\end{figure}

\section{RELATED WORK}
Uncertainties have been addressed with various MCTS approaches in the literature. Kernel Regression UCT \cite{Yee2016} tackles the execution uncertainty in continuous action spaces by employing the Upper Confidence Bound for Trees (UCT) \cite{Kocsis2006} and selecting actions according to a modified upper confidence bound value that incorporates all action assessments by kernel regression. Furthermore, the estimated value of a selected action is refined by applying progressive widening to add similar actions to the search tree. Another concept is \textit{determinization} which describes the process of sampling several deterministic problems with perfect information from a stochastic problem with imperfect information, solving these problems, and fusing their results to get a solution for the original problem \cite{Browne2012}. For instance, Cou\"{e}toux et al. \cite{Couetoux2011} employ UCT with "Double Progressive Widening" in a setting with stochastic state transitions and continuous action and state spaces. Their method expands the set of available actions and the set of sampled outcomes iteratively.
Another example is HOP-UCT \cite{Bjarnason2009} which uses "Hindsight Optimization" (HOP). Several deterministic UCT search trees are constructed, and their results are averaged to determine the action assessments. Ensemble-Sparse-UCT \cite{Bjarnason2009} creates multiples trees and restricts the number of outcomes for an action to a sampling width parameter. Afterwards, the results of the search trees are combined.
A different approach is Information Set MCTS \cite{Cowling2012}, which uses nodes representing information sets. An information set comprises all indistinguishable states for an agent. The approach uses a determinization to restrict the search tree to regions consistent with this determinization.
The real-state uncertainty can be modeled by a POMDP approach \cite{Silver2010a} that uses a search tree with history nodes and an unweighted particle filter representing the belief state. A black box simulator samples successor states and observations, and actions are selected according to the UCT criterion. Generally, POMDPs can be solved online with search trees of belief states with Branch-and-Bound Pruning, Monte Carlo Sampling, or Heuristic Search \cite{Ross2008}.

In the domain of automated driving, the "Toolkit for approximating and Adapting POMDP solutions In Realtime" (TAPIR) \cite{Klimenko2014} based on Adaptive Belief Trees \cite{Kurniawati2016} is successfully applied if the action space is sufficiently small \cite{Hubmann2017, Hubmann2018, Schoerner2019}. Another approach to account for uncertainty is to use Distributional Reinforcement Learning in combination with risk metrics \cite{Bernhard2019}. Instead of learning a return for each state and action, Distributional Reinforcement Learning learns a return distribution. This can be done offline, exposing the agent to various uncertainties during training. Online, during inference, a risk metric is applied to the (risk-neutral) return distribution to select the best action.

\section{BACKGROUND}
The following subsections provide related fundamentals.

\subsection{Monte Carlo Tree Search}
Monte Carlo Tree Search is a Reinforcement Learning based method to search for an optimal decision within an MDP \cite{Browne2012}. It has proven to be applicable to problems with large branching factors \cite{Silver2010a, Silver2016}.
The optimal decision maximizes the (expected) sum of discounted future rewards,
\begin{align}
	\label{eq:basics:return}
	\returnRVar_t &= \rewardRVar_{t} + \discountRate \rewardRVar_{t+1} + \discountRate^2 \rewardRVar_{t+2} + \dots
	= \sum_{k=0}^{\infty} \discountRate^k \rewardRVar_{t+k},
\end{align}
with $\rewardFun$ being the reward function and $\discountRate$ being the discount factor.
It approximates the action-value function through Monte Carlo sampling \cite{Sutton2018} by the arithmetic average of the returns \eqref{eq:basics:return},
\begin{equation}
	\actionValueRVar(\state, \action) = \overline{\returnRVar}^{(\state, \action)} = \frac{1}{\visitCount(\state, \action)} \sum_{j=1}^{\visitCount(\state, \action)} \returnRVar^{(\state, \action)}_{j},
\end{equation}
where $\visitCount(\state, \action)$ denotes the visit count of action $\action$ from state $\state$ and $\returnRVar^{(\state, \action)}_{j}$ is the $j$-th return sample of executing action $\action$ from state $\state$.

The objective is to find the optimal policy $\policy_*$ that provides the maximum expected return. Given the optimal action-value function $\actionValueFun_*(\state, \action) = \max_{\policy}\actionValueFun_{\policy}(\state, \action)$, a deterministic optimal policy can be determined by choosing the action that maximizes $\actionValueFun_{*}(\state, \action')$.

To estimate the optimal action-value function, the MCTS applies the following four steps repeatedly until a computational budget is exhausted \cite{Browne2012}:
\subsubsection{Selection} The selection phase starts from the root node and descends the search tree (e.g., by maximizing the UCT value \eqref{eq:uct} of the next node) until a node is reached that is non-terminal and has not been fully expanded (i.e., not all available actions have been explored). This node is selected.
\subsubsection{Expansion}The expansion phase selects an action (e.g., at random) from the action space and expands the previously selected node by executing the action, generating a new child node.
\subsubsection{Simulation}During the simulation phase, a simulation is run from the new child node according to a rollout policy $\policyRoll$ until a terminal state is reached.
\subsubsection{Update}The update phase traverses the branch of the tree starting from the added child node to the root node and updates the node statistics (i.e., visit count and return of actions).

The selection and expansion process is defined by a tree policy $\policyTree$. For instance, 
the Upper Confidence Bound for Trees (UCT) \cite{Kocsis2006} is calculated during the selection phase for all possible actions $\action$ from a given node $\node$  with
\begin{equation}
	\label{eq:uct}
	\operatorname{UCT}(\node, \action) = \overline{\returnRVar}^{(\node, \action)} + 2 C_p \sqrt{\frac{\log \visitCount(\node)}{\visitCount(\node, \action)}}
\end{equation}
where $C_p \in \realNumbers_{\geq 0}$ is a constant, $\visitCount(\node, \action)$ the visit count of action $\action$ from node $\node$ and $\visitCount(\node)$ is the total visit count of node $\node$.
The child node with the highest UCT value gets selected. UCT solves the exploration-exploitation dilemma by balancing both phases throughout the search, with the first term fostering exploitation and the latter exploration.
Further in-depth information about MCTS can be found in \cite{Browne2012}.

\subsection{Kernel Regression}
\label{sec:kr}

The general objective of regression is to find the regression function
\begin{align}
	m(x) &= \expectedValue[\myRandomVar{Y} | \myRandomVar{X} = x] = \frac{\int_{-\infty}^{\infty} y f(x,y) dy}{\int_{-\infty}^{\infty} f(x,y) dy}
\end{align}
specifying the conditional expected value of a random variable $\myRandomVar{Y}$ given the realization of a random variable $\myRandomVar{X}$ \cite{Watson1964}. Since the joint probability density function $f$ is unknown, \textit{kernel regression} \cite{Nadaraya1964, Watson1964} estimates $f$ by 
\begin{align}
	\hat{f}(x,y) = \frac{1}{n} \sum_{i=1}^{n} \widetilde{\kernel}(x - x_i, y - y_i)
\end{align}
with a kernel $\widetilde{\kernel}$ (i.e., a non-negative smoothing function whose integral over both dimensions equals one) from a finite set of data samples $\{(x_1, y_1), \dots, (x_n, y_n)\}$. Under mild requirements, the estimated regression function can then be formulated as
\begin{equation}
	\hat{m}(x) = \frac{\sum_{i=1}^{n} y_i \widetilde{\kernel}(x - x_i)}{\sum_{i=1}^{n} \widetilde{\kernel}(x - x_i)} = \frac{\sum_{i=1}^{n} y_i \kernel(x, x_i)}{\sum_{i=1}^{n} \kernel(x, x_i)}
\end{equation}
where $\widetilde{\kernel}(\cdot)$ is the "marginal" kernel and the kernel value $\kernel(x, x_i)$ is just a simplified notation for $\widetilde{\kernel}(x - x_i)$ \cite{Watson1964}.

\subsection{Risk Metrics}
\label{sec:riskMetrics}

A risk metric is a "measure" for risk, but it is not a metric in a mathematical sense since it does not represent a distance function. Let $\costRVar: \sampleSpace \rightarrow \realNumbers$ be a random variable assigning costs (in monetary units) that are caused by an action to each outcome $\sample$ of the sample space $\sampleSpace$. Furthermore, let $\mathcal{Z}$ denote the set of all cost random variables, then a risk metric $\riskMetric: \mathcal{Z} \rightarrow \realNumbers$ maps each cost random variable $\costRVar$ to a real number that represents the "amount" of risk \cite{Majumdar2017}. "Good" risk metrics should fulfill the following axioms \cite{Majumdar2017}: \textit{A1. Monotonicity}, \textit{A2. Translation invariance}, \textit{A3. Positive homogeneity}, \textit{A4. Subadditivity}, \textit{A5. Comonotone additivity}, and \textit{A6. Law invariance}.
The proper mathematical definitions can be found in \cite{Majumdar2017}. The following subsections describe two possible risk metrics, the Kernel Regression Lower Confidence Bound (KRLCB) and the Conditional Value at Risk (CVaR). 

\subsubsection{Kernel Regression Lower Confidence Bound (KRLCB)}

The concept of kernel regression (see section \ref{sec:kr}) can be applied to estimate an action value. The kernel regression value
\begin{equation}
	\label{eq:kr}
	\kernelReg[\action | \state, \actionSpace] = \frac{\sum_{\actionB \in \actionSpace} \kernel(\action, \actionB) \actionValueRVar(\state, \actionB) \visitCount(\state, \actionB)}{\sum_{\actionB \in \actionSpace} \kernel(\action, \actionB) \visitCount(\state, \actionB)}
\end{equation}
for an action $\action$ combines the action value estimates $\actionValueRVar(\state, \actionB)$ of all actions $\actionB$ of a finite action set $\actionSpace$ from a state $\state$ weighted by the visit counts $\visitCount(\state, \actionB)$ and a kernel $\kernel$ that specifies the similarity between the actions  \cite{Yee2016}. The denominator in \eqref{eq:kr} is called the kernel density
\begin{equation}
	\label{eq:density}
	\density[\action | \state, \actionSpace] = \sum_{\actionB \in \actionSpace} \kernel(\action, \actionB) \visitCount(\state, \actionB)
\end{equation}
and specifies the exploration of action $\action$ and similar actions. The kernel regression lower confidence bound
\begin{equation}
	\label{eq:krlcb}
	\krlcb[\action | \state, \actionSpace] = \kernelReg[\action | \state, \actionSpace] - c \sqrt{\frac{\log \sum_{\actionB \in \actionSpace} \density[\actionB | \state, \actionSpace]}{\density[\action | \state, \actionSpace]}}
\end{equation}
subtracts a normalized exploration term scaled by a constant $c \in \realNumbers_{\geq 0}$ from the $\kernelReg$ value \cite{Yee2016} and hence penalizes poorly explored actions. In this context, the KRLCB is applied to a return distribution represented by the value estimates $\actionValueRVar(\state, \actionB)$. It can be shown that some of the risk metric axioms in section \ref{sec:riskMetrics} are satisfied if the KRLCB definition is adjusted for a cost distribution.

\subsubsection{Conditional Value at Risk (CVaR)}
\label{sec:riskMetrics:CVaR}

Let $\costRVar$ be a cost random variable. Then, the Value at Risk
\begin{align}
	\label{eq:VaR}
	\VaR_{\alpha}(\costRVar) &:= \min \left\{ z \in \realNumbers \ | \ \probMeasure(\costRVar > z) \leq \alpha \right\} \\
	&= \min \left\{ z \in \realNumbers \ | \ \probMeasure(\costRVar \leq z) \geq 1 - \alpha \right\} \label{eq:VaR:quantile}
\end{align}
specifies the smallest $(1 - \alpha)$-quantile of $\costRVar$ for a given probability $\alpha$ \cite{Rockafellar2002, Majumdar2017}. The Conditional Value at Risk
\begin{equation}
	\label{eq:CVaR}
	\CVaR_{\alpha}(\costRVar) := \expectedValue\left[ Z \ | \ Z \geq \VaR_{\alpha}(\costRVar) \right]
\end{equation}
is defined as the conditional expected value of $\costRVar$ given all costs are greater than or equal to $\VaR_{\alpha}(\costRVar)$. CVaR is a risk metric that satisfies all six axioms \cite{Majumdar2017}. In addition, the Upper Value at Risk
\begin{align}
	\label{eq:VaR+}
	\VaR_{\alpha}^+ (\costRVar) &:= \inf \left\{ z \in \realNumbers \ | \ \probMeasure(\costRVar > z) < \alpha \right\} \\
	&= \inf \left\{ z \in \realNumbers \ | \ \probMeasure(\costRVar \leq z) > 1 - \alpha \right\}
\end{align}
specifies the largest $(1 - \alpha)$-quantile of $\costRVar$ for a given probability $\alpha$
\cite{Rockafellar2002}. $\VaR_{\alpha}^+$ and $\VaR_{\alpha}$ only differ if the cumulative distribution function is constant around $(1 - \alpha)$.

\section{PROBLEM STATEMENT}
The problem of cooperative trajectory planning in scenarios with multiple traffic participants and the associated sources of uncertainty is represented by a multi-agent POMDP \cite{Kaelbling1998}, specified by a tuple $\langle \agents, \states, \actionSpace, \stateTransitionFun,\rewardFun, \observationSpace, \observationFun, \discountRateVec \rangle$ with:
\begin{itemize}
	\item $\agents = \{1, \dots, \numAgents\}$: the set of agents
	\item $\states$: the state space
	\item $\actionSpace = \actionSpace^1 \times \dots \times \actionSpace^\numAgents$: the joint action space including the action space $\actionSpace^i$ for each agent $i \in \agents$. 
	\item $\stateTransitionFun: \states \times \actionSpace \times \states \rightarrow [0,1]$: the state-transition function that specifies the probability $\stateTransitionFun(\state, \actionVec, \state')$ of reaching state $\state'$ after the joint action $\actionVec$ is executed in state $\state$.
	\item $\rewardFun: \states \times \actionSpace \times \states \rightarrow \realNumbers^\numAgents$: the reward function that specifies the joint reward $\rewardFun(\state, \actionVec, \state')$ as a result of the transition from state $\state$ to state $\state'$ by the joint action $\actionVec$.
	\item $\observationSpace = \observationSpace^1 \times \dots \times \observationSpace^\numAgents$: the joint observation space containing the observation space $\observationSpace^i$ for each agent $i \in \agents$.
	\item $\observationFun: \states \times \actionSpace \times \observationSpace \rightarrow [0,1]$: the joint observation function that specifies the probability $\observationFun(\state', \actionVec, \observationVec)$ of making the joint observation $\observationVec$ on the condition that the agents executed the joint action $\actionVec$ and transitioned to the state $\state'$.
	\item $\discountRateVec$: a vector of discount factors. The discount factor $\discountRate^i \in [0,1)$ for an agent $i \in \agents$ accounts for the inherent uncertainty of future rewards by reducing their weight in the return calculation.
\end{itemize}

In this paper, we focus on the real-state uncertainty modeled by the observation function $\observationFun$. The unknown intentions of other traffic participants are explicitly modeled, and the execution uncertainty represented by the state-transition function $\stateTransitionFun$ is addressed by a \textit{"similarity update"} in our previous work \cite{Kurzer2018b}. The objective is to find the optimal joint action $\actionVec$ provided a specific observation $\observationVec$.

\section{APPROACH}
Our approach fosters cooperation between agents by employing a reward function for each agent that also incorporates the rewards for all other agents \cite{Kurzer2018b}. Each agent chooses actions ($\Delta$velocity, $\Delta$lateral position) independently in a continuous action space to achieve the desired velocity and lane while preventing collisions and invalid trajectories. Based on the chosen actions, trajectories are generated using fifth-order polynomials in a Frenet coordinate system \cite{Kurzer2018b, Werling2010}.

We model the accuracy of a sensor system with Gaussian distributions. The accuracy is described by its \textit{trueness} and its \textit{precision} according to ISO 5725-1 \cite{ISO5725-1:1994}. The \textit{trueness} states the deviation of the arithmetic mean of data points from their real value. Since we assume that the sensor systems are unbiased, we set the means of the Gaussian distributions to the real (simulated) values. The \textit{precision} describes the dispersion of the data points and can hence be expressed by the standard deviations of the Gaussian distributions.

The following observation features for the agents' vehicles are modeled as stochastic: longitudinal and lateral position, longitudinal and lateral velocity, length, width, and heading. Furthermore, observations for the longitudinal and lateral position, length, width, and heading of obstacles are stochastic, as well as the observed lane width.

A stochastic observation is represented by an m-dimensional random vector $\mathbf{X} = (\myRandomVar{X}_1, \dots,\allowbreak \myRandomVar{X}_m )^T$ that follows a Gaussian distribution $\mathcal{N}(\bm{\mu}_{\state'}, \mathbf{\Sigma})$ with the mean vector $\bm{\mu}_{\state'}$ and the covariance matrix $\mathbf{\Sigma}$. The elements of the mean vector $\bm{\mu}_{\state'}$ correspond to the real features of the current state $\state'$ since the sensor systems are assumed to be unbiased.
For reasons of simplicity, we assume that the random vector $\mathbf{X}$ contains only mutually independent components. This leads to $\mathbf{\Sigma}$ being a diagonal matrix and each observation feature $\myRandomVar{X}_k$ following its individual Gaussian distribution $\mathcal{N}(\mu_{\state'_k}, \sigma_k^2)$.

Our approach addresses the uncertainty surrounding sensor measurements by combining the results of search trees from different start states, see Fig.~\ref{fig:overview}. These start states $\startStates$ are sampled in a modified way according to the initial belief state $\belief_0$ (i.e., based on the measurement uncertainties). Instead of updating the belief state over time steps within the MCTS, we employ determinization and treat the resulting states as deterministic, applying standard MDP logic. Given the distribution over possible states through the initial belief state and the results from the determinization, more robust actions can be found.
Our approach does not address the update procedure of the belief state over planning cycles. Instead, we use a belief state that solely depends on the current observation and the accuracy of the sensor systems.

The general concept is described in Alg.~\ref{alg:uncertaintyHandling}. 

\begin{algorithm}
	\caption{Uncertainty Handling Concept}
	\label{alg:uncertaintyHandling}
	\begin{algorithmic}[1]
		\Require{belief state $\belief_0$}
		\Ensure{best action vector $\bestActionVec$}
		\State create initial start states $\startStates$
		\For{iteration $i \gets 1,\dots,\numIterations$}
		\If{start states $\startStates$ shall be expanded}
		\State $\startState \gets$ create new start state
		\State $\startStates \gets \startStates \cup \{\startState\}$
		\Else
		\State $\startState \gets$ select existing start state $\in \startStates$
		\EndIf
		\Statex
		\State $\node \gets$ select node in tree of $\startState$ to be expanded
		\State $\node' \gets$ expand $\node$ by creating new node according to selected action vector
		\State add $\node'$ as child node of $\node$
		\State run simulation from $\node'$
		\State update tree of $\startState$ according to simulation results
		\State update start state $\startState$
		\EndFor
		\Statex
		\State $\bestActionVec \gets$ \finalSelPol{} selects best action vector given search trees
	\end{algorithmic}
\end{algorithm}

\subsection{Start State Policies}
The following describes the selection and expansion policies for start states.

\subsubsection{Selection}\label{sec:selection}
Since the belief about the real state of the environment is continuous and an MCTS-based approach can only handle a finite set of discrete states, we apply \textit{progressive widening} (cf.~\cite{Coulom2007, Chaslot2008}) to extend the set of start states $\startStates$ iteratively.
During each iteration, it is checked whether
\begin{equation}
	\label{eq:startStates:progWidening}
	\left| \startStates \right| \geq c_{pw} \visitCount^{\alpha_{pw}} 
\end{equation}
with the start states $\startStates$, the constants $c_{pw} \in \realNumbers_{\geq 0}$, and $\alpha_{pw} \in [0,1)$ and the iteration count $\visitCount$ holds true. If it does not, the start states are expanded by adding a new start state generated according to section \ref{sec:expansion}.
Otherwise, a start state is selected uniformly at random from the set of collision-free and valid start states $\startStates' \subseteq \startStates$. In summary, \eqref{eq:startStates:progWidening} specifies the number of sampled start states for a given number of iterations.

\subsubsection{Expansion}\label{sec:expansion}
Given the current belief, a new start state is generated according to a specific sampling process. Since we model the belief as Gaussian distributions, we also use Gaussian distributions to sample start states.
Let $\mathbf{X} = \left(\myRandomVar{X}_1, \dots,  \myRandomVar{X}_m \right)^T$ be an m-dimensional random vector that is normally distributed ($\mathbf{X} \sim \mathcal{N}(\bm{\mu}, c \mathbf{\Sigma})$) with the mean vector $\bm{\mu}$ and the covariance matrix $\mathbf{\Sigma}$ scaled by a factor $c$. Each component $\myRandomVar{X}_k : \sampleSpace \rightarrow \realNumbers$ specifies the value of a feature, for instance the lateral position of a vehicle. The mean vector $\bm{\mu}$ and the covariance matrix $\mathbf{\Sigma}$ are set to the corresponding parameters of the current belief state.
Then, a start state is sampled according to this distribution $\mathcal{N}(\bm{\mu}, c \mathbf{\Sigma})$. If the sampled start state is invalid or in collision, the factor $c$ gets adjusted, and a new attempt is conducted. This procedure is repeated until the start state meets the conditions or a maximum limit is reached, with $c$ defined as
\begin{equation}
	\label{eq:sigmaFactor}
	c = \min \left\{ \left( c_{step} \right)^{l_{step}} , c_{max}\right\},
\end{equation}
where $c_{step} \in \realNumbers_{\geq 0}$ is a constant, $l_{step}$ is a step counter and $c_{max} \in \realNumbers_{\geq 0}$ is the maximally allowed value. The step counter
\begin{equation}
	\label{eq:sigmaStep}
	l_{step} = \left\lfloor \frac{l_{attempt}}{l_{stepSize}} \right\rfloor
\end{equation}
is dependent on the number of total attempts conducted so far $l_{attempt} \in \realNumbers_{\geq 0}$, and the step size $l_{stepSize} \in \realNumbers_{\geq 0}$ specifies the number of attempts for one step.

The mentioned procedure is executed independently for the creation of several start states. Thus, the $i$-th start state is a realization of the random vector $\mathbf{X}_i \sim \mathcal{N}(\bm{\mu}, c_i \mathbf{\Sigma})$ with an individual factor $c_i$.

If $c_i$ increases, the variance of sampled start states also increases, and hence the probability of obtaining a collision-free and valid start state rises. Furthermore, the usage of $c_i > 1$ fosters the finding of more robust actions, as they have to perform well from a greater variety of start states. On the other hand, the constant $c_{max}$ limits the scaling of the variance to a realistic level. Future work may examine whether evaluations from specific start states (e.g., near collisions and invalidities) should be treated differently.

\subsection{Final Selection}
\label{sec:finalSelection}

A final selection policy selects the most robust action vector given the constructed search trees. We present two different variants, namely the Lower Confidence Bound and the Conditional Value at Risk. Both variants follow the same steps stated in Alg.~\ref{alg:uncertaintyHandling:finalSelection}. For each agent $i \in \agents$, a set of action candidates $\actionCandidates^i$ is determined, comprising all actions whose assessment contribute to the final result. This set only contains actions explored from valid and collision-free start states with a visit count exceeding a lower threshold. Since we assume that planning is only conducted from valid and collision-free states, results from other start states that cannot represent the actual state of the environment are rejected. As the variance of action value estimates decreases with increasing visit counts, the threshold can be seen as a variance reduction measure. If the condition is met for an action, this action is appended to the action candidates, and the start state from which the action was evaluated is also stored. Therefore, the set of action candidates is not a mathematical set because it is possible that a specific action, e.g., "accelerate", is included several times from different start states. On the other hand, a specific action from a specific start state is only added once. In this sense, all elements of $\actionCandidates^i$ are distinct.

In case of an empty set of action candidates for an agent, the default action of maintaining the current velocity and the lateral position is selected. Otherwise, kernel regression is conducted to combine the action assessments. In general, the density of an action candidate $\action$ is defined as
\begin{equation}
	\label{eq:density:actionCandidates}
	\density[\action | \actionCandidates^i] = \sum_{\actionB \in \actionCandidates^i} \kernel(\action, \actionB) \visitCount(\state_{\actionB}, \actionB)
\end{equation}
with kernel $\kernel$ and the visit count $\visitCount(\state_{\actionB}, \actionB)$ of action candidate $\actionB$ from the corresponding start state $\state_{\actionB}$. We use the kernel specified by the Gaussian radial basis function
\begin{equation}
	\label{eq:kernel}
	\kernel(\action, \action') = \exp \left( -\gamma \left\lVert \action - \action' \right\rVert^2 \right)	
\end{equation}
as a similarity measure for two actions $\action$ and $\actionB$. Consequently, the density is a measure for the exploration of an action $\action$ and similar actions to $\action$.
The kernel regression value is calculated as
\begin{equation}
	\label{eq:kr:actionCandidates}
	\kernelReg[\action | \actionCandidates^i] = \frac{\sum_{\actionB \in \actionCandidates^i} \kernel(\action, \actionB) \actionValueRVar(\state_{\actionB}, \actionB) \visitCount(\state_{\actionB}, \actionB)}{\sum_{\actionB \in \actionCandidates^i} \kernel(\action, \actionB) \visitCount(\state_{\actionB}, \actionB)}
\end{equation}
where $\actionValueRVar(\state_{\actionB}, \actionB)$ is the estimated action value of action candidate $\actionB$ evaluated from the corresponding start state $\state_{\actionB}$. Thus, it is a sum of value estimates weighted by the visit counts and the similarity between $\action$ and the actions candidates.
After "action candidate values" $\actionCandidateValues^i$, describing the final assessment of the action candidates, are calculated according to the policy variants introduced in the next subsections, the action with the maximum action candidate value is chosen for the respective agent.

\begin{algorithm}[tbp]
	\caption{Basic Final Selection Policy}
	\label{alg:uncertaintyHandling:finalSelection}
	\begin{algorithmic}[1]
		\Require search trees of the planning phase
		\Ensure best action vector $\bestActionVec$	
		\State $\bestActionVec \gets \langle\rangle$
		\Comment{empty best action vector}
		
		\ForAll{agent $i \in \agents$}
		\State $\actionCandidates^i \gets$ \Call{getActionCandidates}{$i$}
		\If{$\actionCandidates^i = \emptyset$}
		\State store default action in $\bestActionVec$
		\Else
		\State $\densityValues^i, \kernelRegValues^i \gets$ \Call{kernelRegression}{$\actionCandidates^i$}
		\State $\actionCandidateValues^i \gets$ \Call{getACValues}{$\actionCandidates^i$, $\densityValues^i$, $\kernelRegValues^i$}
		\State $\bestAction \gets \argmax_{\action \in \actionCandidates^i} \actionCandidateValues^i (\action)$
		\State store $\bestAction$ in $\bestActionVec$
		\EndIf
		\EndFor
		\State \Return $\bestActionVec$
	\end{algorithmic}
\end{algorithm}

\subsubsection{Kernel Regression Lower Confidence Bound (KRLCB)}

This policy variant employs the Kernel Regression Lower Confidence Bound (KRLCB) \cite{Yee2016} to combine the action assessments from the search trees. First, the function \textproc{kernelRegression} determines the density $\density[\action | \actionCandidates^i]$ and kernel regression value $\kernelReg[\action | \actionCandidates^i]$ for all action candidates $\action \in \actionCandidates^i$ according to \eqref{eq:density:actionCandidates} and \eqref{eq:kr:actionCandidates}, respectively. Then, the function \textproc{getACValues} calculates the KRLCB value
\begin{equation}
	\label{eq:krlcb:actionCandidates}
	\krlcb[\action | \actionCandidates^i] = \kernelReg[\action | \actionCandidates^i] - c \sqrt{\frac{\log \sum_{\actionB \in \actionCandidates^i} \density[\actionB | \actionCandidates^i]}{\density[\action | \actionCandidates^i]}}
\end{equation}
given a constant $c \in \realNumbers_{\geq 0}$ and the precomputed $\density[\action | \actionCandidates^i]$ and $\kernelReg[\action | \actionCandidates^i]$. The second term in \eqref{eq:krlcb:actionCandidates} represents the normalized exploration of $\action$ and similar action candidates. By subtracting this term, rarely explored action candidates and thus poor value estimates are penalized, leading to a lower confidence bound value. Since the KRLCB value incorporates the assessment from all constructed trees, a well-explored action that performs well from many start states has a large KRLCB value and hence is selected finally as a robust action.

\subsubsection{Conditional Value at Risk (CVaR)}

\begin{figure}[tbp]
		\centering
		\def\svgwidth{0.7\columnwidth}
		\footnotesize
		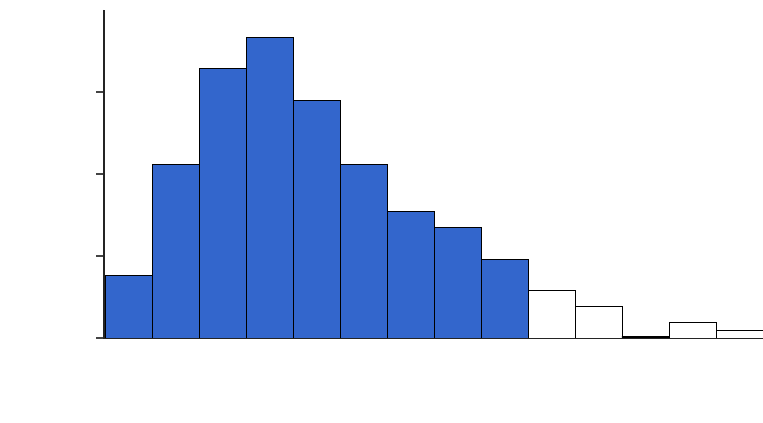
		\caption{CCVaR and upper VaR for an exemplary probability density function (p.d.f.) describing a return distribution.}
		\label{fig:CVaR:return:hist}
\end{figure}

This policy variant employs the Conditional Value at Risk (CVaR) \cite{Rockafellar2002, Majumdar2017}. In section \ref{sec:riskMetrics:CVaR}, CVaR was introduced for a cost random variable $\costRVar$. Since we operate in the context of MDPs and POMDPs, we derive a consistent CVaR definition for a reward random variable $\rewardRVar := - \costRVar$. Given \eqref{eq:VaR}, the Value at Risk of $\costRVar$ can be transformed with
\begin{equation}
	\label{eq:VaR:Z->R}
	\begin{aligned}
		\VaR_{\alpha}(\costRVar) &= \min \left\{ z \in \realNumbers \ | \ \probMeasure(- \rewardRVar > z) \leq \alpha \right\} \\
		& = - \max \left\{ r \in \realNumbers \ | \ \probMeasure(\rewardRVar < r) \leq \alpha \right\} \\
		& = - \inf \left\{ r \in \realNumbers \ | \ \probMeasure(\rewardRVar \leq r) > \alpha \right\} \\
		& = - \inf \left\{ r \in \realNumbers \ | \ \probMeasure(\rewardRVar > r) < 1 - \alpha \right\} \\
		& \overset{\eqref{eq:VaR+}}{=} - \VaR_{ 1 - \alpha}^+ (\rewardRVar)
	\end{aligned}
\end{equation}
to the negative Upper Value at Risk of $\rewardRVar$. Given a cost random variable $\costRVar$, the best action $\bestAction$ of a finite set $\actionSetAgent$ of available actions for agent $i$ has the smallest CVaR value \eqref{eq:CVaR:bestAction} and hence the smallest risk. The equation manipulations
\begin{align}
	\bestAction &= \argmin_{\action \in \actionSetAgent} \CVaR_{\alpha}(\costRVar) \label{eq:CVaR:bestAction} \\
	&= \argmin_{\action \in \actionSetAgent} \expectedValue \left[ - \rewardRVar \ | \ -\rewardRVar \geq \VaR_{\alpha}(\costRVar) \right] \nonumber \\
	&\overset{(\ref{eq:VaR:Z->R})}{=} \argmin_{\action \in \actionSetAgent} \expectedValue \left[ - \rewardRVar \ | \ \rewardRVar \leq \VaR_{1 - \alpha}^+ (\rewardRVar) \right] \nonumber \\
	&= \argmax_{\action \in \actionSetAgent} \expectedValue \left[ \rewardRVar \ | \ \rewardRVar \leq \VaR_{1 - \alpha}^+ (\rewardRVar) \right] \label{eq:CVaR:bestAction:R}
\end{align}
show that the best action $\bestAction$ can also be selected by maximizing an expected value based on the reward random variable $\rewardRVar$ \eqref{eq:CVaR:bestAction:R}. We denote this expected value as "Complementary Conditional Value at Risk" (CCVaR) and define it as
\begin{equation}
	\label{eq:CVaR:reward}
	\CCVaR_{1-\alpha}(\rewardRVar) := \expectedValue\left[ \rewardRVar \ | \ \rewardRVar \leq \VaR_{1 - \alpha}^+ (\rewardRVar) \right] \text{.}
\end{equation}
In the context of MDPs and POMDPs, an agent selects the action with the largest $\CCVaR_{1-\alpha}(\returnRVar)$ value where $\returnRVar$ denotes the return \eqref{eq:basics:return}. CCVaR and the upper VaR for an exemplary return distribution are illustrated in Fig.~\ref{fig:CVaR:return:hist}.

The policy variant using the CCVaR, which is described in Alg.~\ref{alg:uncertaintyHandling:finalSelection:CVaR}, must determine a return distribution for each action candidate $\action \in \actionCandidates^i$ at first. This is achieved by calculating kernel regression values for different start states as "unweighted particles" representing the distribution. The function \textproc{kernelRegression} determines for each action candidate $\action \in \actionCandidates^i$ and for each start state $\startState \in \startStates$ the action candidates $\actionCandidates^i_{\startState} \subseteq \actionCandidates^i$ that were evaluated from $\startState$. The density $\densityValues^i_{\action, \startState} := \density[\action | \actionCandidates^i_{\startState}]$ \eqref{eq:density:actionCandidates} describes the exploration of $\action$ and similar action candidates in the tree of $\startState$. Only if this density $\densityValues^i_{\action, \startState}$ is greater than or equal to a threshold $w_{min} \in \realNumbers_{\geq 0}$, a "particle" for the start state $\startState$ is added to the return distribution. This condition ensures lower variance of the return estimates. If the condition holds, the kernel regression value $\kernelRegValues^i_{\action, \startState} := \kernelReg[\action | \actionCandidates^i_{\startState}]$ \eqref{eq:kr:actionCandidates} is calculated and the results $\densityValues^i_{\action, \startState}$ and $\kernelRegValues^i_{\action, \startState}$ are appended to the sets $\densityValues^i$ and $\kernelRegValues^i$, respectively. After the kernel regression procedure is completed, the return distribution for an action candidate $\action \in \actionCandidates^i$ is represented by the unweighted particle set of kernel regression values
\begin{equation}
	\kernelRegValues^i_{\action} = \left\{\kernelRegValues^i_{\action, \startState} \ | \ \startState \in \startStates \wedge \kernelRegValues^i_{\action, \startState} \in \kernelRegValues^i \right\} \subseteq \kernelRegValues^i \text{.}
\end{equation}
In the next step, the function \textproc{getACValue} checks whether $\kernelRegValues^i_{\action}$ has enough elements for a meaningful representation. Let
\begin{equation}
	\label{eq:numOfRelStartStates}
	\begin{split}
	N_{\startState} = | \{\actionCandidates^i_{\startState} \ &| \ \startState \in \startStates \\
	&\wedge \actionCandidates^i_{\startState} = \left\{ \text{action candidates from } \startState \right\} \subseteq \actionCandidates^i \\
	&\wedge \actionCandidates^i_{\startState} \neq \emptyset \} |
	\end{split}
\end{equation}
be the number of start states from which actions have been appended to the action candidates $\actionCandidates^i$. If the final selection follows section \ref{sec:finalSelection} for the initialization of $\actionCandidates^i$, $N_{\startState}$ corresponds to the number of collision-free and valid start states visited more often than a threshold. Then, the condition
\begin{equation}
	\label{eq:condSigDist}
	|\kernelRegValues^i_{\action}| \geq c_{m} N_{\startState}
\end{equation}
with a constant $ c_{m} \in [0,1]$ indicates whether $\kernelRegValues^i_{\action}$ is meaningful. If \eqref{eq:condSigDist} is satisfied, the $\CCVaR_{1-\alpha}(\kernelRegValues^i_{\action})$ \eqref{eq:CVaR:reward} with probability $\alpha$ is used as the action candidate value for $\action \in \actionCandidates^i$. Otherwise, the action candidate value is set to negative infinity to avoid the selection of an action that has been poorly assessed due to the sparse set $\kernelRegValues^i_{\action}$.

The CCVaR approach favors robust actions since it combines the performances from several start states and limits the influence of high-return outliers from few specific start states. 

\begin{algorithm}[tbp]
	\caption{CVaR Final Selection Policy}
	\label{alg:uncertaintyHandling:finalSelection:CVaR}
	\begin{algorithmic}[1]
	    \Require{action candidates $\actionCandidates^i$ for agent $i$}
		\Ensure{density distributions $\densityValues^i$ and kernel regression value distributions $\kernelRegValues^i$ for agent $i$}
		\Function{kernelRegression}{$\actionCandidates^i$}
		\State $\densityValues^i \gets \emptyset$
		\State $\kernelRegValues^i \gets \emptyset$
		
		\ForAll{action $\action \in \actionCandidates^i$}
		\ForAll{start state $\startState \in \startStates$}
		\State $\actionCandidates^i_{\startState} \gets \left\{ \text{action candidates from } \startState \right\} \subseteq \actionCandidates^i$
		\State $\densityValues^i_{\action, \startState} \gets \density[\action | \actionCandidates^i_{\startState}]$
		\Comment{see \eqref{eq:density:actionCandidates}}
		\If{$\densityValues^i_{\action, \startState} \geq w_{min}$} \label{alg:uncertaintyHandling:finalSelection:CVaR:wmin}
		\State $\kernelRegValues^i_{\action, \startState} \gets \kernelReg[\action | \actionCandidates^i_{\startState}]$ 
		\Comment{see \eqref{eq:kr:actionCandidates}}
		\State append $\densityValues^i_{\action, \startState}$ to $\densityValues^i$ and $\kernelRegValues^i_{\action, \startState}$ to $\kernelRegValues^i$
		\EndIf
		\EndFor
		\EndFor
		\State \Return $\densityValues^i$, $\kernelRegValues^i$
		\EndFunction
		\Statex
		
		\Require{action candidates $\actionCandidates^i$, densities $\densityValues^i$, kernel regression values $\kernelRegValues^i$ for agent $i$}
		\Ensure{CCVaR values $\actionCandidateValues^i$ for agent $i$}
		\Function{getACValues}{$\actionCandidates^i$, $\densityValues^i$, $\kernelRegValues^i$}
		\State $\actionCandidateValues^i \gets \emptyset$
		\LComment{number of start states from which actions have been appended to action candidates}
		\State $N_{\startState} \gets$ see \eqref{eq:numOfRelStartStates}
		
		\ForAll{action $\action \in \actionCandidates^i$}
		\State $\kernelRegValues^i_{\action} \gets \left\{\kernelRegValues^i_{\action, \startState} \ | \ \startState \in \startStates \wedge \kernelRegValues^i_{\action, \startState} \in \kernelRegValues^i \right\}$ 
		
		\If{$|\kernelRegValues^i_{\action}| \geq c_{m} N_{\startState}$ with $c_{m} \in [0,1]$}
		\State $\actionCandidateValues^i_\action \gets \CCVaR_{1-\alpha}(\kernelRegValues^i_{\action})$ 
		\Comment{see \eqref{eq:CVaR:reward}}
		\Else
		\State $\actionCandidateValues^i_\action \gets - \infty$ 
		\EndIf
		\State append $\actionCandidateValues^i_\action$ to $\actionCandidateValues^i$
		\EndFor
		\State \Return $\actionCandidateValues^i$
		\EndFunction	
	\end{algorithmic}
\end{algorithm}

\section{Experiments}
\begin{figure*}[tbp]
	\centering
	\begin{subfigure}[b]{0.666\columnwidth}		
		\centering
		\def\svgwidth{\columnwidth}
		\scriptsize
		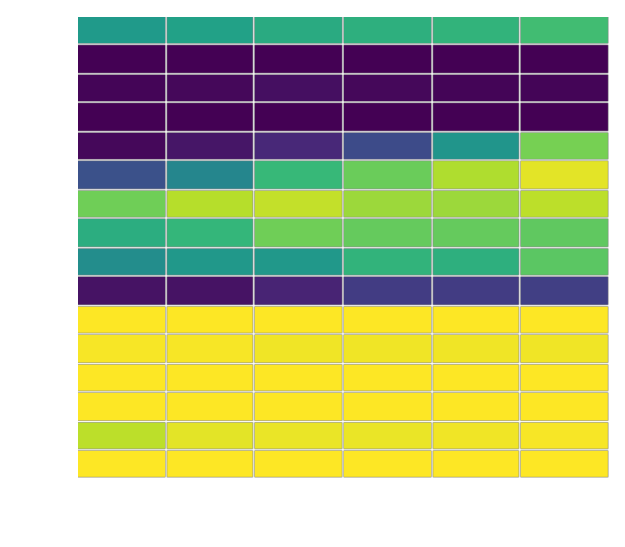
		\caption{baseline deterministic environment}
		\label{fig:results:noise_off_handling_off}
	\end{subfigure}
	\hfill
	\begin{subfigure}[b]{0.666\columnwidth}
		\centering
		\def\svgwidth{\columnwidth}
		\scriptsize
		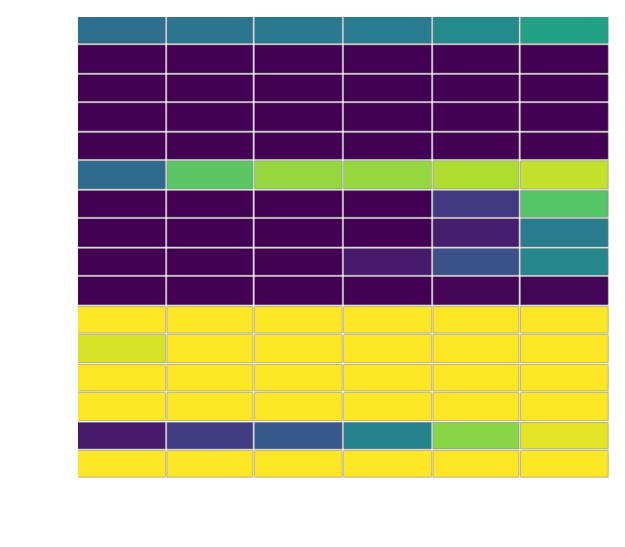
		\caption{KRLCB deterministic environment}
		\label{fig:results:krlcb_noise_off_handling_on_depth}
				
	\end{subfigure}
	\hfill
	\begin{subfigure}[b]{0.666\columnwidth}
		\centering
		\def\svgwidth{\columnwidth}
		\scriptsize
		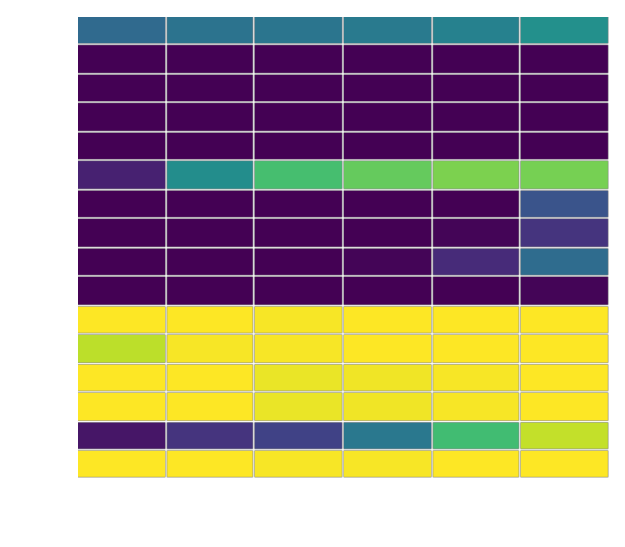
		\caption{CVaR deterministic environment}
		\label{fig:results:dist_noise_off_handling_on_depth}
	\end{subfigure}
	\begin{subfigure}[b]{0.666\columnwidth}
		\centering
		\def\svgwidth{\columnwidth}
		\scriptsize
		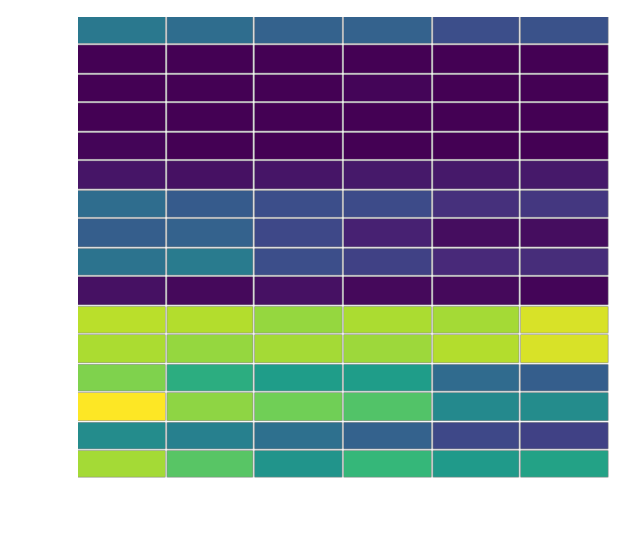
		\caption{baseline uncertain environment}
		\label{fig:results:noise_on_handling_off}
	\end{subfigure}
	\hfill
	\begin{subfigure}[b]{0.666\columnwidth}
		\centering
		\def\svgwidth{\columnwidth}
		\scriptsize
		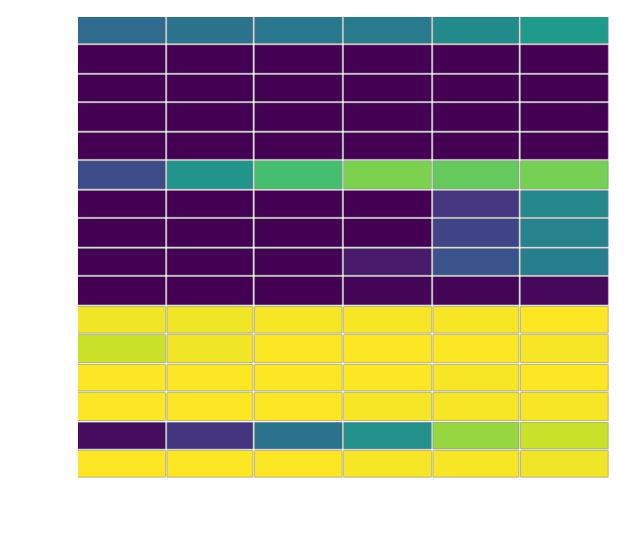
		\caption{KRLCB uncertain environment}
		\label{fig:results:krlcb_noise_on_handling_on_depth}
	\end{subfigure}
	\hfill
	\begin{subfigure}[b]{0.666\columnwidth}
		\centering
		\def\svgwidth{\columnwidth}
		\scriptsize
		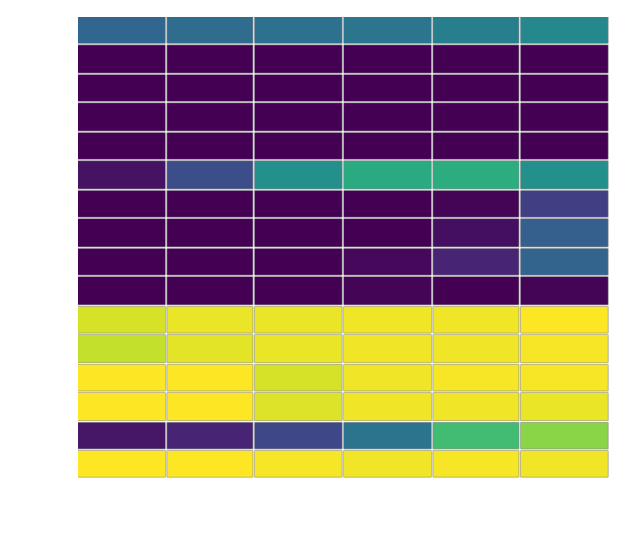
		\caption{CVaR uncertain environment}
		\label{fig:results:dist_noise_on_handling_on_depth}
	\end{subfigure}
	\caption{Comparison of the success rates between the baseline, KRLCB, and CVAR final selection policies in deterministic (a, b, c) and uncertain (d, e, f) scenarios (SC) for different numbers of iterations.}
	\label{fig:results}
\end{figure*}

We evaluated our approach using 15 different scenarios representing merging, overtaking, and bottleneck situations as well as obstacle mazes with up to 8 vehicles \cite{Kurzer2020}. The scenarios, which are illustrated online\footnote{\url{https://url.kurzer.de/ProSeCo-Scenarios}}, describe various urban situations. A scenario is completed when each agent has left the scenario's region of interest. Each experiment was conducted with 300 random seed values with our simulator. The results were assessed with the success rate indicating the fraction of collision-free and valid trajectories. An agent's trajectory is valid if the agent stays within the road boundaries and if the trajectory is physically drivable. We compare the results for different numbers of iterations $\numIterations$ (cf.~Alg.~\ref{alg:uncertaintyHandling}). The baseline, which is the cooperative trajectory planner of our previous work \cite{Kurzer2018b}, provides the results of Fig.~\ref{fig:results:noise_off_handling_off} in a deterministic environment (i.e., without uncertain measurements). Scenarios one to six with two or three agents yield a success rate close to \SI{100}{\percent}. For scenarios eight to eleven, the success rates increase with the number of iterations as expected due to the improved action value estimates and the exploration of new actions. Scenarios 13 to 15, which describe more complex situations such as obstacle mazes, cannot be solved with 8000 iterations or less since more iterations are necessary for a proper exploration of the action spaces and for a greater planning depth into the future. Thus, the results for these scenarios are neglected in the evaluation. 


If the baseline approach is applied in an uncertain environment with noisy observations, the results worsen according to Fig.~\ref{fig:results:noise_on_handling_off}. The success rates do not improve significantly or even deteriorate in some scenarios with increasing iterations. This is reasonable since the action value estimates overfit to the observed state that can differ from the true state, possibly leading to the selection of worse actions.


Our final selection policy with the KRLCB enhances the results significantly (see Fig.~\ref{fig:results:krlcb_noise_on_handling_on_depth}). The success rates almost reach \SI{100}{\percent} in scenarios one to six. Furthermore, the performance increases with increasing iterations. This is sensible since better estimates from more trees are combined, robustifying action selection. If our approach is applied in a deterministic environment (see Fig.~\ref{fig:results:krlcb_noise_off_handling_on_depth}), the results are similar but do not achieve the performance of the baseline. This is reasonable, as the modeling and consideration of non-existent uncertainties leads to the selection of suboptimal actions. Especially in more complex scenarios such as SC07-SC10, more iterations are necessary to mitigate this effect.

%

Similarly, the CVaR final selection policy also improves the results in an uncertain environment significantly (see Fig.~\ref{fig:results:dist_noise_on_handling_on_depth}). The results are close to the results of the KRLCB policy variant, but more iterations are necessary for the same success rates. Moreover, the same statements also pertain in a deterministic environment (see Fig.~\ref{fig:results:dist_noise_on_handling_on_depth}).



The average runtime per planning step for the baseline and our KRLCB and CVaR variants are stated in table \ref{tab:runtime}. A planning step provides a sequence of actions for a maximum duration of \SI{10}{\second}, but only the first action is executed for \SI{0.8}{\second} in our simulation, leading to the next planning step given a new observation. Both variants yield similar overheads compared to the baseline. For instance, their durations are greater by $\approx \SI{7}{\percent}$ at 1000 iterations, \SI{19}{\percent} at 2000 iterations, and \SI{6}{\percent} at 8000 iterations compared to the baseline. The large overhead at 2000 iterations is due to the chosen parameters for progressive widening of the set of start states.

\begin{table}[tbp]
    \centering
    \caption{Average runtime per planning step in milliseconds}
    \label{tab:runtime}
    \renewcommand{\arraystretch}{1.2}
    \begin{tabular}{l|c|c|c|c|c|c}
        \multirow{2}{*}{Approach} & \multicolumn{6}{c}{Number of iterations} \\ \cline{2-7}
                 & 250 & 500 & 1000 & 2000 & 4000 & 8000 \\ \hline
        Baseline & 25 & 52 & 112 & 249 & 638 & 1381 \\
        KRLCB    & 27 & 53 & 120 & 299 & 695 & 1464 \\
        CVaR     & 27 & 52 & 121 & 296 & 696 & 1479
    \end{tabular}
\end{table}

\section{CONCLUSION}
We propose an MCTS-based approach to plan robust cooperative trajectories in an uncertain environment represented by a multi-agent POMDP. Our approach samples start states in a modified way according to an initial belief state, constructs deterministic search trees for all start states, and combines their results by applying risk metrics to the return distributions created by kernel regression. We employ the Kernel Regression Lower Confidence Bound (KRLCB) and the Conditional Value at Risk (CVaR) risk metrics to finally select robust actions that perform well from many start states. The results demonstrate that our approach outperforms our prior cooperative trajectory planner significantly in uncertain environments due to noisy observations while incurring only a slight runtime overhead.




\section*{ACKNOWLEDGMENT}
We wish to thank the German Research Foundation (DFG) for funding the project Cooperatively Interacting Automobiles (CoInCar), within which the research leading to this contribution was conducted. The information and views presented in this publication are solely the ones expressed by the authors.

\bibliographystyle{IEEEtran}
\bibliography{new_library,library}

\end{document}